\documentclass[mathleft
]{an}
\usepackage{graphicx}
\usepackage{times}
\begin{document}

\Pagespan{789}{}
\Yearpublication{2006}%
\Yearsubmission{2005}%
\Month{11}%
\Volume{999}%
\Issue{88}%

\title{Where do the Progenitors of Millisecond Pulsars come from?}

\author{Ali Taani\inst{1}\fnmsep\thanks{Corresponding author:
  \email{alitaani@bao.ac.cn}\newline}
\and  Chengmin Zhang\inst{1} \and Mashhoor Al-Wardat \inst{2}
\and
Yongheng Zhao \inst{1}}
\titlerunning{Where do the Progenitors of Millisecond Pulsars come from?}
\authorrunning{Taani et al.}
\institute{ National Astronomical Observatories, Chinese Academy of
Sciences, Beijing 100012, China
 \and Department of Physics, Al-Hussein Bin Talal University, P.O.Box 20, 71111, Ma'an, Jordan}

\received{30 May 2005}
\accepted{11 Nov 2005}
\publonline{later}


\abstract{%
  Observations of a large population of Millisecond Pulsars (MSPs) show
a wide divergence in the orbital periods (from approximately hours
to a few months). In the standard view, Low-Mass X-Ray Binaries
(LMXBs) are considered as progenitors for some MSPs during the
recycling process. We present a systematic study that combines
different types of compact objects in binaries such as Cataclysmic
Variables (CVs), LMXBs and MSPs. We plot them together in the so
called Corbet diagram. Larger and different samples are
needed to better constrain the result as a function of the
environment and formations. A scale diagram showing the
distribution of MSPs for different orbital periods and the aspects
for their progenitors relying on Accretion Induced Collapse (AIC) of
white dwarfs in binaries. Thus massive CVs ($M \geq 1.1
M_{\odot}$) can play a vital role on binary evolution, as well as of
the physical processes involved in the formation and evolution of
neutron stars and their magnetic fields, and could turn into binary
MSPs with different scales of
 orbital periods; this effect can be explained by the AIC process.
This scenario also suggests that some fraction of isolated MSPs in
the Galactic disk could be formed through the same channel,
formingthe contribution of some CVs to the single-degenerate
progenitors of Type Ia supernova. Furthermore, we have refined the
statistical distribution and evolution by using updated data. This
implies that the significant studies of compact objects in binary
systems can benefit from the Corbet diagram.}

\maketitle

\section{Introduction}
For many years, studies of the population of millisecond pulsars
(MSPs) have been plagued by small-number statistics in order to
obtain a more qualitative inspection of their progenitors. In a
binary system the mass of the Neutron Star (NS) is larger than the
mass of its companion (Bhattacharya \& van den Heuvel 1991), which
causes it to have a short period (a few hours to days). However, it
should be noted that 20\% of MSPs are isolated (Camilo et al. 2001),
but it is not clear how they have lost their presumed past
companions. A widely held view is that the isolated cases are
results of the primary's Super Nova (SN) type Ia explosions, which
would have caused  the loss of their companions by either
evaporation or kicking out of the system (Bhattacharya 1996;
Ferrario \& Wickramasinghe 2007ab).

Several statistical studies have questioned whether the known
low-mass X-ray binary (LMXB) populations could produce the observed
MSPs (e.g. Alpar et al. 1982; Srinivasan \& van den Heuvel 1982;
Burderi et al. 2007; Liu et al. 2011 ). In the accepted
scenario of this evolution, an old and quiescent NS is spun up to
millisecond periods by the accretion of matter from its companion,
during the LMXB phase of evolution. All of the previously
mentioned papers, however, focused on systems in short orbital
periods.

Moreover, Accretion Induced Collapse (AIC) of white dwarfs
(WDs) has been proposed as an alternative source of recycled pulsars
sufficient to obviate the difficulties with the standard model (van
den Heuvel 2004; Hurley et al. 2010).

While other authors have questioned the viability of an AIC
origin for the recycled pulsars on theoretical and statistical
grounds (van den
 Romani 1990; Heuvel 2004). Ferrario \& Wickramasinghe (2007ab)
argued that the AIC channel can form binary MSPs of all of
observable types. Champion et al. (2008) proposed that the AIC could
produce the observed orbital parameters of PSR J1903+0327. This
scenario has a number of attractive features and it is desirable to
have quantitative tests of expected results. This study aimed to
investigate the binary MSPs and their progenitors within different
binaries with compact companions from different catalogues.
Furthermore we are paying particular attention to the AIC process in
the observed sample of CVs.

However, there are several numerical simulation models that
employ different physical mechanisms to model an AIC process.
Dessart et al. (2007) have performed 2D multigroup
neutrino-transport MHD simulations. Also Abdikamalov et al. (2010)
have presented results from an extensive set of general-relativistic
AIC simulations using a microphysical finite-temperature equation of
state during collapse; they investigated a set of 114 progenitor
models in axisymmetric rotational equilibrium, with a wide range of
rotational configurations, temperatures and central densities, and
resulting WD masses. In addition, Liu et al. (2011) have studied the
Corbet diagram with regard to Be/X-ray binaries, and they
investigated the progenitors of IGR J18483-0311 and IGR J11215-5952.
Yan et al. (2010) studied the correlation between LMXBs and normal
pulsars in the Corbet diagram.

Progress towards the current sample led to a resurgence of
discoveries and interest from observers and theorists concerning the
new distribution of MSPs. The mechanisms responsible for their
formation in Globular Clusters and \\Galactic disks are different
(Sutantyo \& Li 2000; Ivanova et al. 2008). Consequently, we show
that from studying the "Corbet-diagram" (correlation of spin period
vs. orbital period diagram) one is able to obtain clues about how
the evolution of MSPs differs from those in binary and isolated
systems (Yan et al. 2010).

This paper is motivated by previous work investigating the
statistical properties of binary MSPs and their progenitors with
long orbital periods (a few months). We defer further discussion on
the model until the data collection is completed. The outline of
this paper is as follows: In section 2, we review the formation
routes including the recycling process and AIC of a white dwarf. In
section 3, we discuss the Corbet diagram. We present the data
acquisition and target selection in sections 4. Summary and
conclusions are presented in section 5.

\begin{figure}
\includegraphics[width=6.5cm, angle=0]{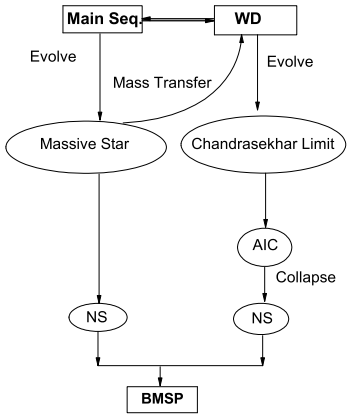}
\caption{The flow-chart illustrates the AIC evolutionary scenario of
MSPs. } \label{fig1}
\end{figure}

\begin{figure}
\includegraphics[width=6.5cm, angle=0] {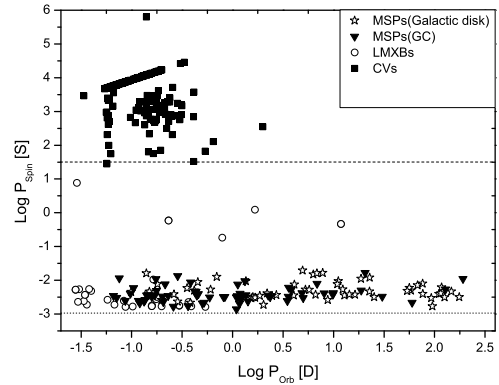}
\caption{The spin period vs. the orbital periods for binaries
(Corbet diagram). Different symbols denote different types of binary
systems. Here the CVs are represented (black squares), LMXBs (white
circles), MSPs in the Galactic disk (white stars) and MSPs in
globular cluster (black triangles). The horizontal dashed line at a
spin period of 30s splits the figure into two parts; WDs with
$P>30s$ and NSs with $P<30s$. Dotted line represents the spin period
at 1 ms.} \label{fig2}
\end{figure}

\section{The Formation Routes for MSPs}
Interaction between the NS and its companion star is clearly
important in determining the activity of any particular system. In
order to understand this interaction and the consequent activity, we
shall first discuss the formation scenarios in which the MSPs are
formed, beginning with the standard route of the recycling process.
On the other hand, we shall also discuss an alternative scenario by
AIC of WDs.

\subsection {Fast-spinning ``Recycling Process"}
 The standard evolutionary scenario in which Binary MSPs are formed
from LMXBs is considered as a powerful tools for the study of
orbital kinematics in binary MSPs systems. The evolutionary path is
briefly described as follows: A high-field ($B \sim
$$10^{12}$-$10^{13}$ G), rapidly rotating NS is born in a binary
with a low-mass (~1$M_{\odot}$) main-sequence companion star. During
the SN that produces the NS, mass loss and kick imparted on the NS
cause the orbit to be eccentric. This NS remains bound to its
companion and spins down like a normal pulsar for the next $10^{6}$
to $10^{7}$ yrs (Lyne et al. 2004; Lorimer 2009), until passing the
so-called ``death line" in the magnetic field-spin period (B–P)
diagram. As the companion evolves, it eventually fills its Roche
lobe. At this stage, transfer of matter from the companion to the NS
creates an X-ray binary and tidal friction serves to circularize the
orbit.

Mass accretion onto the NS gives rise to X-ray emission and induced
magnetic field decay (the mechanisms for the field decay induced by
accretion are, however, not well understood), therefore it spins up
to a millisecond period. As a result, the systems can be observed as
LMXBs, which generate  brighter shorter lived sources. (Bhattacharya
\& van den Heuvel 1991). When the companion loses almost all of its
envelope and mass transfer ceases, the endpoint of the evolution is
a circular binary with a weak field MSP NS and an He or CO WD, the
remaining core of the companion. A fast-growing population of
eccentric (e $> 0.5$) PSR J1748-2021B. Binary MSPs have been
recently revealed in globular clusters (Freire et al. 2007). Note
that the formation of an LMXB invokes a binary evolution in which
the progenitor system has to lose ~90\% of its initial mass and up
to ~99\% of its initial orbital angular momentum (van den Heuvel
2009).

\begin{table*}
\centering
 \caption{\bf List of some Massive CVs.$^{*}$}
\setlength{\tabcolsep}{10pt}

\centering \label{table1}
\begin{tabular}{lcccccl}
\hline \hline
Name  &   $M_{\odot}$ &  $P_{orb}$  &    $M_{1}/M_{2}$   \\

\hline
CAL 83  &   1.3 &   1.047 &   --    \\
CI Aql  &   1.2 &   0.618 &   0.8    \\
BV Cen  &   1.24    &   0.611 &   1.12   \\
RU Peg  &   1.21    &   0.374  &   1.29     \\
MU Cen  &   1.2 &   0.342   &   1.2        \\
BF Eri  &   1.28    &   0.270 &   2.3       \\
VY Scl  &   1.22    &   0.232  &   2.7        \\
BD Pav  &   1.15    &   0.179  &   1.5         \\
U  Gem  &   1.2 &   0.176 &   2.6      \\
CU Vel  &   1.23    &   0.078  &   5.0       \\
DP Leo  &   1.2 &   0.062 &   5.5       \\
\textbf{IP Peg}  & 1.16 &  0.16 & 2.08 \\
\textbf{RX And}  & 1.14    & 0.209 & 2.4\\
\textbf{EY Cyg} & 1.1 &  0.49 & 2.27 \\
\textbf{WW Hor} & 1.1 &   0.08 & 5.8 \\
\textbf{SS Aur}  & 1.08    & 0.182 & 2.8\\
\hline

\multicolumn{4}{l}{$^{*}$The data are taken from Ritter \& Kolb
(2011).}
\end{tabular}
\medskip
\end{table*}

\begin{table*}
\centering
 \caption{\bf Descriptive statistical data for histograms of orbital period.}
\setlength{\tabcolsep}{10pt}

\centering \label{table1}
\begin{tabular}{lccccccl}
\hline \hline
Name  &   Total number &  Mean  & Stand. Dev. & Sum & Mini & Median & Max. \\

\hline
CVs  &   178 &   -0.916 &  0.255 & -163.11 & -1.267 & -0.894 & 0.30    \\
LMXBs  &   36 &  -0.835 &  0.655 & -30.054 & -1.555 & -0.80 & 1.073    \\
MSP (Galactic disk)  &  91 &   -0.356 &  0.242 & -214.39 & -2.807 & -2.405 & -1.713    \\
MSP (GC)  &   64 &   -0.053 &  0.822 & -3.45 & -1.179 & -0.159 &2.282    \\
\hline

\multicolumn{4}{l}{}
\end{tabular}
\medskip
\end{table*}

\begin{table*}
\centering
 \caption{\bf Descriptive statistical data for histograms of spin period .}
\setlength{\tabcolsep}{10pt}

\centering \label{table1}
\begin{tabular}{lccccccl}
\hline \hline
Name  &   Total number &  Mean  & Stand. Dev. & Sum & Mini & Median & Max. \\

\hline
CVs  &   120 &   3.53 &  0.527 & 621.133 & 1.755 & 3.718 & 4.458    \\
LMXBs  &   37 &  -1.978 &  1.215 & -73.201 & -2.79 & -2.497 & 2.057   \\
MSP (Galactic disk)  &  91 &   -2.356 &  0.242 & -214.39 & -2.807 & -2.405 & -1.713    \\
MSP (GC)  &   120 &   -2.37 &  0.21 & -283.39 & -2.86 & -2.38 &-1.76   \\
\hline

\multicolumn{4}{l}{}
\end{tabular}
\medskip
\end{table*}

\subsection {Accretion Induced Collapse of  white dwarfs}
Despite there being a clear-cut theoretical prediction of Accretion
Induced Collapse (AIC) to form MSPs (Canal \& Schatzman 1967; Nomoto
1984, 1987), only recently have transient surveys had sufficient
depth field of view to potentially detect and characterize these
events through the  optical and infrared light curves and spectra
for the Nickel-rich outflows (Metzger et al. 2009ab \& and Darbha et
al. 2010). However not all compositions can be involved in AIC. A
key factor in the choice is the density of the chemical composition
of the system (Isern \& Hernanz 1994).

In C-O WDs, the high temperature and great density can  produced
ignition. Consequently, the thermonuclear explosion causes the
nuclear burning shell to be ejected leading to disruption of the WD
and explosion as an SN Ia with a rate $10^{-4}$ yr$^{-1}$ (Popov \&
Prokhorov 2007). This event can help us to possibly measure
distances and  the cosmological constant accurately (Bradely 2010).
However, CVs may have implications for the still open
question of which objects actually produce SNe Ia, owing to their
long formation times and their relatively low accretion rates
(Zortovic et al. 2010).

The scenario for an O-Ne-Mg WD is begin once the nuclear reaction
starts at the center, the burning propagates throughout the entire
star, then the ignition can happen in the interior. The burning
which propagates outward makes central temperatures and pressures
high enough to lead to a collapse of the WD, and induces the
contraction of the star. This continues until it reaches the
Chandrasekhar limit, and finally the star collapses homogeneously
leading to the formation of an NS. Consequently, the AIC can be a
significant contribution to the total number and evolution of LMXB
(van den Heuvel 1997); notice that the minimum density for obtaining
a collapse is ${\rm5.5 \times 10^{9} g/cm^3}$.

Depending on the strength of the WD's magnetic field, the matter
flowing through the Lagrangian point can form either a full or
partial accretion disk or else it will follow the magnetic field
lines down to the surface at the magnetic poles (Warner 1995). It is
therefore important to monitor long-period CV systems in order to
gain insight into the dynamics and binary evolution in those systems
most likely to undergo AIC.

At present, it is a well established fact that low mass WD stars
should be formed during the evolution in close binary systems. Here
the orbital evolution of these binaries, and the mass-transfer rate
is driven by magnetic braking of the secondary for long-period
systems, as well as by gravitational radiation for short-period
systems (Beuermann \& Riensch 2002). This situation is in agreement
with the possibility to apply the AIC route in binaries to produce
binary MSPs. It occurs either because the WD began accreting as an
ONeMg or because the accretion rate onto the WD does not allow them
to remain a C-O WD. The AIC formation channel of MSP evolution is
depicted in Fig. ~\ref{fig1}.

\section {The Corbet diagram}
Various aspects of the statistical work performed by distributions
for orbital-period and spin-period elucidate the origin and the
mechanisms responsible for MSP formation. The Corbet diagram
refers to the $P_{spin}-P_{orb}$ relation, proposed first by Corbet
(1984, 1985), who noted there is a $P_{spin} \alpha P^{2} _{orb}$
correlation for Be systems, whereas $P_{spin}\alpha P^{4/7} _{orb}$
for systems with radially expanding winds was soon identified by van
den Heuvel \& Rappaport (1987) (Norton et al. 2004; Yan et al. 2010). Fig.~\ref{fig2}
shows the four types of binaries, which represent a panoramic
perspective of binaries. The LMXBs (white circles) have short
orbital and spin periods. It can be seen clearly that the CVs (black
squares) are almost all spread at the top left corner, with
 orbital periods that typically range from approximately 0.06 to
2 days, and long spin periods (slow rotators). That means in the
case of a massive CV, if it accretes  enough matter from the
companion till it  reaches the Chandrasekhar limit,  it will
collapse to become an NS during the AIC  producing a short period
binary MSP.

It should be noticed that at the moment no model is able to reproduce all of
the observed binary properties; in particular, the predicted
distributions of mass, orbital period and mass ratio tend not to match the
observations very well (Reggiani \& Meyer 2011). If a model can be
described by the correlation between these objects, many more CVs
would be contributed as good progenitors of MSPs. This can be
readily shown via a Venn diagram in Fig. 3, which shows how four
sets can intersect. However observations of long orbital periods,
more massive CVs and LMXBs in disparate environments can at least
put constraints on theoretical binary formation models (Norton et al. 2004).

 Our list of CVs is drawn from the new catalogue of Ritter \& Kolb (2011), as a starting point. The number of CVs shows that a few sources are
massive $\sim M \geq 1.1 M_{\odot}$  (see Table ~\ref{table1}).
Fig. 4 shows the Gaussian distrbution of massive CVs, with
maximum at $M_{WD}\sim 1.3M\odot$, where the mean mass of these
sources is $M_{WD} = 1.19 \pm 0.15M\odot$. On the other hand, long
orbital periods in the MSP distribution can be elucidated by the
significant imparted kicks to the companion during the SN Ia. The
kick speed, $V_{kick}$, is modelled by a Maxwellian distribution, as
given by Hansen \& Phinney (1997) with a dispersion $V\sigma$ = 190
Kms$^{-1}$

\begin{equation}
P(V_{kick})=\sqrt{\frac{2}{\pi}}\frac{V_{kick}^{2}}{V_{\sigma}^{3}}e^{-\frac{V_{kick}^{2}}{2V_{\sigma}^{2}}}
\end{equation}

 \begin{figure}
\includegraphics[width=6.5cm, angle=0]{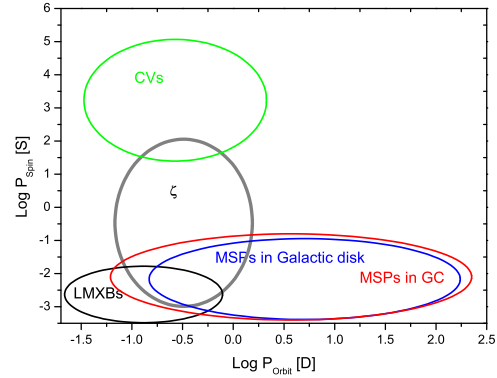}
\caption{A Venn diagram showing the relation among four different
kinds of objects: MSPs in Galactic disk, MSPs in globular cluster,
LMXBs and CVs. $\zeta$ represents the model.}. \label{fig3}
\end{figure}

 \begin{figure}
\includegraphics[width=6.5cm, angle=0]{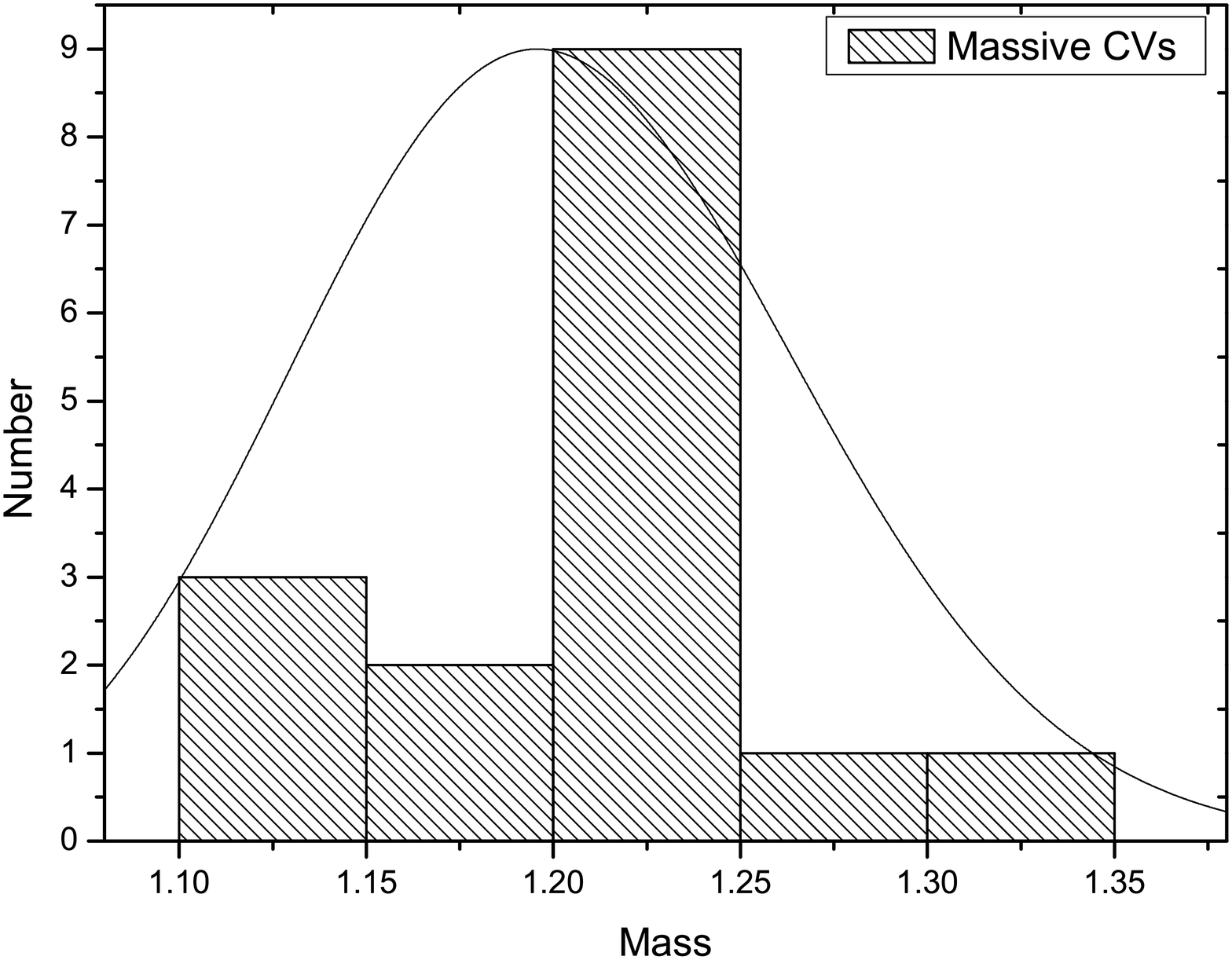}
\caption{Mass distribution of the observed sample of massive CVs.
The solid line is the curve fitted using a Gaussian function. Data
from Ritter \& Kolb (2011) were used to construct this distribution.
} \label{fig4}
\end{figure}


 \begin{figure}
\includegraphics[width=6.5cm, angle=0]{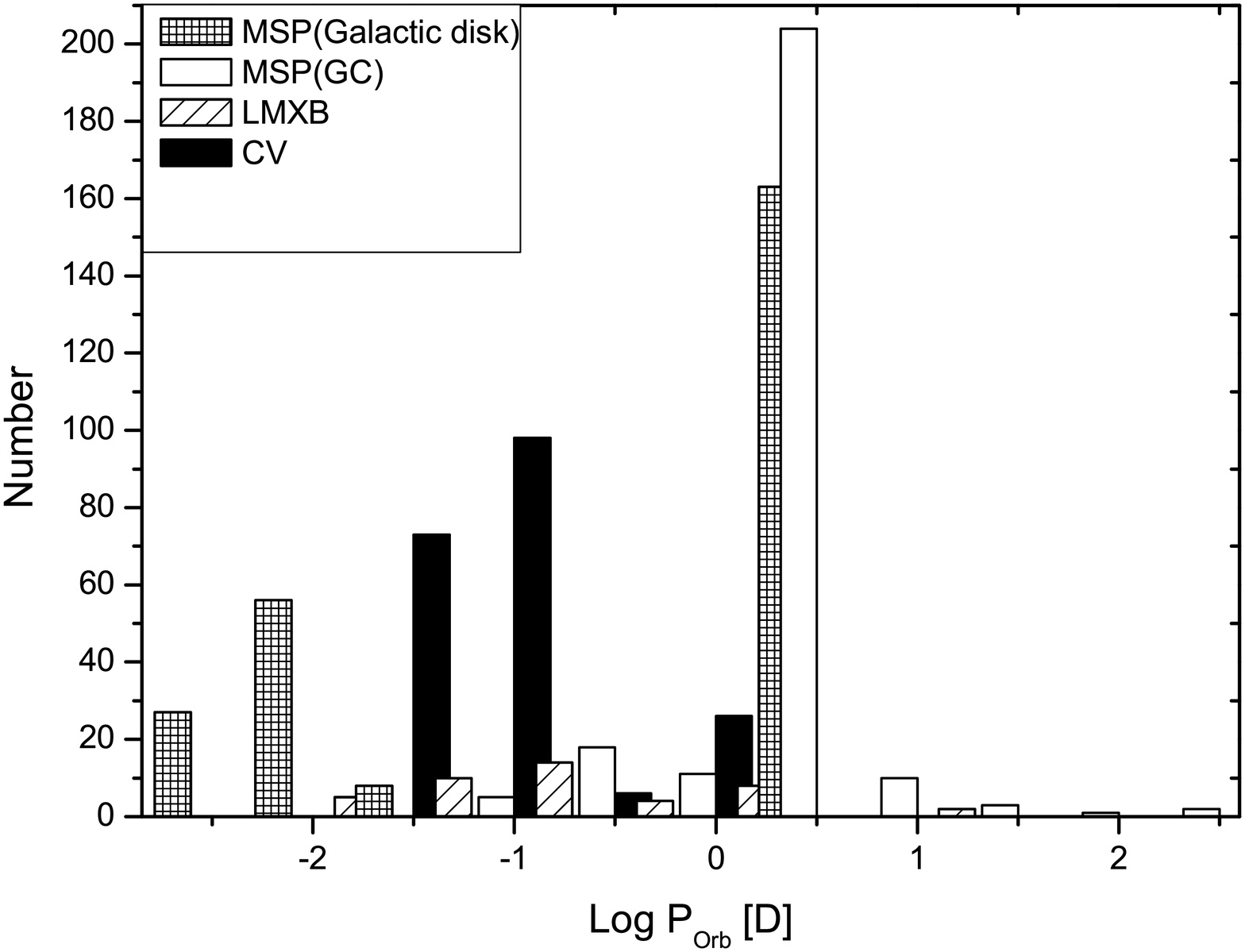}
\caption{The distribution of orbital-periods for four different
kinds of objects: MSPs in Galactic disk (dense histogram), MSPs in
globular cluster (solid), LMXBs (sparse) and CVs ( black).}
\label{fig5}
\end{figure}

 \begin{figure}
\includegraphics[width=7.5cm, angle=0]{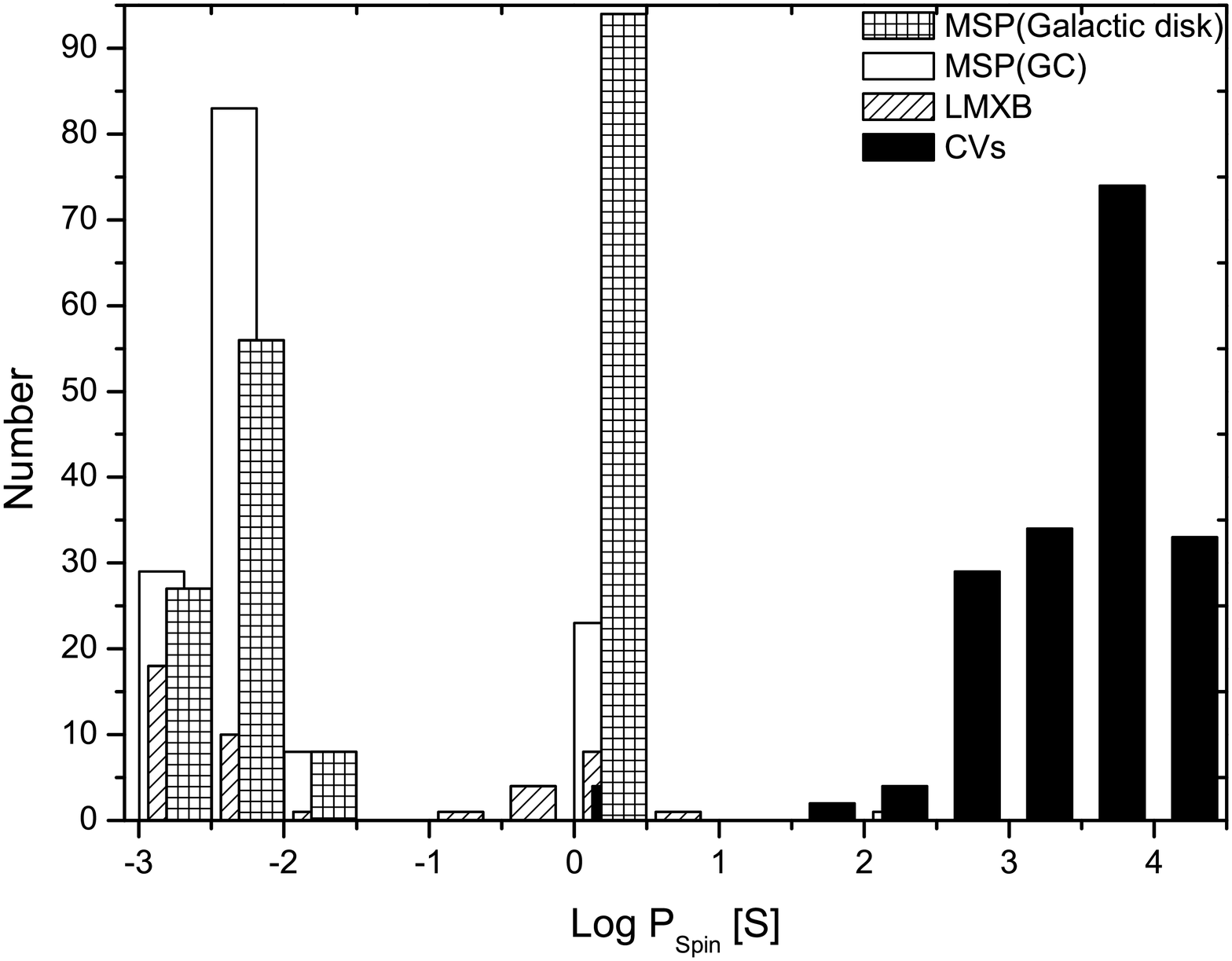}
\caption{The distribution of spin-periods for four different kinds
of objects: MSPs in Galactic disk (dense histogram), MSPs in
globular cluster (solid), LMXBs (sparse) and CVs ( black).}
\label{fig6}
\end{figure}

\begin{table*}
\caption{ Parameters of binary systems with compact companions.}
\label{tlab}

\setlength{\tabcolsep}{1.5pt}
\renewcommand{\arraystretch}{0.9}

\label{table2}
\begin{tabular}{lccccccl}

\hline \hline \noalign{\smallskip}
Group & Sub-group & Obser. evid. & $P_{orb}$ &  $P_{s}$ & $M_{c}$ &  $B$  & Ref.\\
  &   &   & (D) & (S) &  $M_{\odot}$ & (G) & \\
\hline \noalign{\smallskip}

High-mass companion &   NS + NS  &   PSR B1913+16 & 0.32
& 0.059 & 1.4398&   $10^{10}$  $^{i}$   &$^{1}$ \\
   &   NS + (ONeMg) WD &   PSR J1435-61 & 1.35 &  0.093 & $< 1.75$&   $5\times10^8$ &$^{2}$\\

     &  NS + (CO) WD &   PSR J1802-21 & 0.7 &  0.012 & 0.9 &  $9.7\times10^8$  &$^{3}$\\
\hline \noalign{\smallskip}

Low-mass companion &   NS + (He) WD &   PSR 0437-47 & 5.74 & 0.005 & 0.2&  $3\times 10^8$ $^{ii}$  &$^{4}$\\
   &     &   PSR J1744-39 & 0.19 &  0.172 &  0.08&   $1.7\times10^{10}$   & $^{5}$\\

\hline \noalign{\smallskip} Non-recycled pulsar &   NS + (CO) WD   &
PSR J0621+10 & 8.32 & 0.028& 0.75 &   $1.2\times10^{9}$ $^{iii}$   &$^{6}$ \\
Un-evolved Companion &   Non-degenerate  &   PSR 1259-63 &  0.047  & 1236.9 & 30 &   $3\times10^{11}$  &$^{7}$\\
\hline \noalign{\smallskip}
Millisecond Pulsars   &  Millisecond companion &   PSR J1903+032 & 0.002 & 0.015& $\sim~0.15-0.45$&  $10^8$ - $10^9$    &$^{8}$ \\
& X-ray transients &   IGR J00291+59 & 0.102 & 0.001 & $<1.6$ &  $3 \times10^8$ $^{iv}$    &$^{9}$ \\
\hline \noalign{\smallskip}
Cataclysmic Variables    &  (CO) WD + (He) WD &   AM Her. & 0.139 & 11960& 0.76&   10-80 $10^8$ $^{v}$   &$^{10}$ \\
\hline \noalign{\smallskip}

\end{tabular}

\medskip

{$^{1}$ Weisberg et al. 2010, $^{2}$ Jacoby et al. 2006,$^{3}$
Ferdman et al. 2010, $^{4}$ Breton et al. 2007,$^{5}$ van Kerkwijk
et al 2005, $^{6}$ Ray et al. 1992, $^{7}$: Liu \& Li 2009,$^{8}$
Marco et al. 1997, $^{9}$ van den Heuvel 2009, $^{10}$ Ritter \&
Kolb 2011.

$^{i}$ Jasinta et al. 2005, $^{ii}$ Zavlin et al. 2002, $^{iii}$
Eric et al. 2002, $^{iv}$ Papitto et al.  2006, $^{v}$  Sion et al.
2006

 \textbf{$Abbreviations$}: \textbf{$P_{s}$}: Spin period in seconds.
\textbf{$P_{orb}$}: Orbital period in days. \textbf{$M_{c}$}: Mass
of companion in solar mass. Ref.: References for the listed
parameters. \textbf{$B$}: Magnetic field in Gauss. Obser. evid.:
Observational evidence for each group.}
\medskip%
\end{table*}

If the binary orbit is disrupted, this leaves a single MSP. This
result agrees with the model by Podsiadlowski et al. (2002) and Taam
(2004). The MSPs occupy a region in this diagram much larger than
that of the LMXBs, and the process concurs with an evolutionary
scenario of a binary MSP.

In order to investigate the distribution of two observed quantities
for the four kinds of astronomical objects more \\clearly, we
produced two histograms (see Figs.~\ref{fig3} \&~\ref{fig4}) for the
distribution of orbital and spin periods. Fig.~\ref{fig3} shows
relatively Gaussian and regular distributions for most objects. A
large fraction of these systems has an orbital period ranging
between 40 minutes to 2 days, except for a few
 MSPs sources in a globular cluster which have relatively longer orbital periods.
It should
be noticed that the data clearly suffer from selection effects, in
particular the lack of many long orbital period systems such as CVs,
that will ultimately determine the answer about the progenitors of
MSPs.

 Concerning the histogram for spin period (Fig.~\ref{fig4}), we see that the
distribution is almost bimodal with spin periods ranging from 30
seconds to 8 hours.  The majority of spin periods for the MSPs
population is in the sub-second region ${\rm\sim 20 ms}$. It
is interesting to note that few systems in globular clusters have
spin periods shorter than those in the Galactic disk. This is may
have due to a different formation mechanisms or shorter life-time to
gravitational decay (Manchester 2006). There is a quite dramatic
cut-off in the spin periods ranging from 7 to 40 seconds; which may
be due to the sensitivity of the radio observation process and the
selection effects. The statistical parameters for orbital
and spin period histograms are listed in Table 2 and 3,
respectively. It is noteworthy to mention here that if we
apply the empirical formula of the Keplerian frequency, this leads
to deriving the value of spin period for the MSP from WD as:

\begin{eqnarray}
\frac{P_{WD}}{P_{NS}}\sim(\frac{R_{WD}}{R_{NS}})^{3/2}\sim10^{4.5}
\end{eqnarray}
where the minimum value of spin period for a WD is\\ $P_{WD}\sim30
s$. Therefore
\begin{eqnarray}
P_{NS}\sim10^{-3} s
\end{eqnarray}
can be considered as a secure range for the NS spin value.
This clearly supports the view that the AIC effect can play an
important role in deriving the spin period for an MSP.

\section{Observational Data and Target Selection}
 We gathered a relatively large number of observations for binary systems with
compact companions from different \\catalogues. The sample contains
a wide variety of sources, such as MSPs,
 LMXBs and CVs. It has two observed quantities: orbital period and spin
  period measurements.

  The collected observations show that the number of \\MSPs listed in the ATNF
  (Australia Telescope National Facility) Pulsar Catalogue (Manchester et al. 2005)\footnote{http://www.atnf.csiro.au/research/pulsar/psrcat/.html}
  is now close to 2000 pulsars; among them we have 211 MSPs, \\which is considered a relatively small
  number. 120 MSP sources are in globular clusters and 91 are in the galactic disk.
  Since MSPs in galactic disk and globular clusters may have
  different properties, we divided the sample into two groups. It
  should be noticed that the measurement for MSPs in the galactic disk is
  relatively accurate and reliable (Manchester 2006).

The data for CVs were distributed as follows:
 \begin{itemize}
    \item[-] 100 objects
from Lilia ferrario/\\(private communications)
\item[-]  34 objects from the Koji Mukai IP
page\footnote{http://asd.gsfc.nasa.gov/Koji.Mukai/iphome/systems/.html},
\item[-] 52 objects were added from (Ritter \&  Kolb 2011), Cataclysmic
Binaries, Low-Mass X-Ray Binaries and Related Objects
\footnote{http://www.mpa-garching.mpg.de/RKcat/}.
\item[-] One IP source, V2069 = IGR J21237 (Martino et al. 2009)
\end{itemize}

 The data of LMXBs were bifurcated into two main catalogues:
\begin{itemize}
    \item  25 objects; Cataclysmic Binaries, Low-Mass X-Ray Binaries and Related
Objects$^{1}$, (Ritter \&  Kolb 2011).
\item  12 objects from the
catalogue of low-mass X-ray binaries in the Galaxy, LMC and SMC
(Fourth edition) (Liu et al. 2007).
\end{itemize}

Table~\ref{table2} summarizes the parameters of some compact
\\companion binary systems with a number of different physical
characteristics.

\section{Summary and Conclusions}
The primary goal of this work is to gather in a single place an
updated list of different compact objects,  to be used for tests
against various theoretical predictions of characteristics of MSPs
progenitors. Several conclusions can potentially be derived from
this work

\begin{enumerate}
    \item  Based upon updated data and a large number of objects,
   the analyzed distribution of the observed sample indicates revisions and
refinement for the evolution of binary MSPs and their progenitors.
This highlights the contribution value of studying the Corbet
diagram (similar to the H-R or B-P diagrams) to understand evolution
and formation mechanism events for different binaries.

\item  The recycling scenario proved our extensive sample for the short
orbital period of such binaries and very small companion masses.

\item  The AIC scenario is potentially significant, and it remains to be
determined whether alternative evolutionary scenarios need to be
invoked. Further work will certainly clarify this issue as
known sources are better characterized.

\item  Our results show that massive CVs increase the proportion of
progenitors for binary MSPs with short orbital periods ($ <  2 $
hours). This result agrees with some theoretical predictions such as
those by Podsiadlowski et al. (2002) and Taam (2004).

\item  Massive CVs predict a significant fraction of progenitors of present day binary
MSP (i.e. CAL 83 and CAL 87). If matter is accreted at a
rate of $M\sim10^{16}gs^{-1}$, and the total mass accreted exceeds a
critical value $\Delta M_{crit}\sim 0.1- 0.2M_{\odot}$, then it will
be recycled to become an MSP after the mass reaches the
Chandrasekhar limit during the AIC process.


\item An alternative scenario that depends on a definite chemical composition, is a
result of an accreting C-O WD; at such low mass transfer
rates, the nuclear burning of the hydrogen accreted on the surface
of the WD takes place in the form of an SN Ia, in which most or all
of the accreted matter is ejected, consequently imparting
significant kicks to the companion creating long orbital periods. In
some cases this process disrupts the system and creates a single
MSP.

\item The long orbital periods occur for half of the MSP population, depending on their
mass ratio (q) according to
\begin{eqnarray}
 q~~\left\{
      \begin{array}{ll}
        M_{c}>M_{MSP}, & \hbox{Short orbit} \\
        M_{c}<M_{MSP}, & \hbox{Long orbit}
      \end{array}
    \right.
\end{eqnarray}

Once mass transfer starts,
the loss of angular momentum causes the orbit to shrink, and thus
the orbital period becomes shorter. In the case of CVs, the mass of
the companion is less than the mass of CVs, therefore, all these
systems appear in a close orbit scale.




\item  The LMXBs appear to have orbital periods shorter than entire MSP
population. This clearly supports the view that LMXB progenitors
must satisfy a large number of evolutionary and structural
constraints of MSPs. On the \\other hand, the LMXBs
population has instigated active discussion about the possibility
that at least some of them originate from AIC. Probably, the process
will make sense in the evolutionary scenario of binary MSPs or other
mechanisms we have not encountered yet. There maybe an otherwise
undetectable fraction of X-ray MSPs since all pulsar surveys have
some limiting flux density. Future work will go steps further, using
the more extensive data set now available from Ritter \& Holb (2011)
and other sources.

\item  Further binary studies and massive CVs and long orbital periods are needed to study the
dependence of the $P_{Spin}-P_{orb}$ relation on dynamical processes
and to test possible variations in the orbital periods, spin periods
and the mass ratio distribution of different formation mechanisms.
Therefore, there is still a debate about how to construct a unified
model to explain the evolution of LMXB, CVs and MSPs. How does the
process of dynamical AIC collapse usually gets started. And there
are related debates over what processes are principally responsible
for isolated MSPs.

Finally, however the current available data for CVs are not
sufficiently accurate or numerous to allow precise analysis. But we
hope that the results of this work will constitute a base for
further studies on the observational properties of massive CVs and
long orbital periods.


\end{enumerate}


\section*{Acknowledgements}
We are grateful for the discussions with Edward van den Heuvel, and
Andrew Norton. The authors would like to thank Lilia Ferrario for AM
Hers data. The research presented here made an extensive use of the
2011 version of the ATNF Pulsar Catalogue (Manchester et al. 2005).
We also thank the anonymous referee for a careful reading of our
manuscript and for numerous useful comments. This research has been
supported by NSFC (No.10773017) and National Basic Research Program
of China (2009CB824800).

\end{document}